\documentclass[aps,twocolumn,prd,nofootinbib,superscriptaddress]{revtex4}
\usepackage{amsmath}
\usepackage{graphicx}
\usepackage{dcolumn}
\usepackage{bm}
\usepackage{amssymb}
\usepackage{latexsym}

\newcommand{\be}{\begin{equation}}
\newcommand{\ee}{\end{equation}}
\newcommand{\bq}{\begin{eqnarray}}
\newcommand{\eq}{\end{eqnarray}}

\begin{document}


\title{Reconstructing generalized ghost condensate model with dynamical dark
energy parametrizations and observational datasets}

\author{Jingfei Zhang}
\affiliation{School of Physics and Optoelectronic Technology, Dalian
University of Technology, Dalian 116024, People's Republic of China}
\author{Xin Zhang}
\affiliation{Institute of Theoretical Physics, Chinese Academy of
Sciences, P.O.Box 2735, Beijing 100080, People's Republic of China}
\affiliation{Interdisciplinary Center of Theoretical Studies,
Chinese Academy of Sciences, P.O.Box 2735, Beijing 100080, People's
Republic of China}
\author{Hongya Liu}
\affiliation{School of Physics and Optoelectronic Technology, Dalian
University of Technology, Dalian 116024, People's Republic of China}

\begin{abstract}
Observations of high-redshift supernovae indicate that the universe
is accelerating at the present stage, and we refer to the cause for
this cosmic acceleration as ``dark energy''. In particular, the
analysis of current data of type Ia supernovae (SNIa), cosmic
large-scale structure (LSS), and the cosmic microwave background
(CMB) anisotropy implies that, with some possibility, the
equation-of-state parameter of dark energy may cross the
cosmological-constant boundary ($w=-1$) during the recent evolution
stage. The model of ``quintom'' has been proposed to describe this
$w=-1$ crossing behavior for dark energy. As a
single-real-scalar-field model of dark energy, the generalized ghost
condensate model provides us with a successful mechanism for
realizing the quintom-like behavior. In this paper, we reconstruct
the generalized ghost condensate model in the light of three forms
of parametrization for dynamical dark energy, with the best-fit
results of up-to-date observational data.

\end{abstract}

\maketitle

\section{Introduction}\label{sec:intr}

It has been confirmed admittedly that our universe is experiencing
an accelerating expansion at the present time, by many cosmological
experiments, such as observations of large scale structure (LSS)
\cite{LSS}, searches for type Ia supernovae (SNIa) \cite{SN}, and
measurements of the cosmic microwave background (CMB) anisotropy
\cite{CMB}. This cosmic acceleration observed strongly supports the
existence of a mysterious exotic matter, dark energy, with large
enough negative pressure, whose energy density has been a dominative
power of the universe. The astrophysical feature of dark energy is
that it remains unclustered at all scales where gravitational
clustering of baryons and nonbaryonic cold dark matter can be seen.
Its gravity effect is shown as a repulsive force so as to make the
expansion of the universe accelerate when its energy density becomes
dominative power of the universe. The combined analysis of
cosmological observations suggests that the universe is spatially
flat, and consists of about $70\%$ dark energy, $30\%$ dust matter
(cold dark matter plus baryons), and negligible radiation. Although
we can affirm that the ultimate fate of the universe is determined
by the feature of dark energy, the nature of dark energy as well as
its cosmological origin remain enigmatic at present. However, we
still can propose some candidates to interpret or describe the
properties of dark energy. The most obvious theoretical candidate of
dark energy is the cosmological constant $\Lambda$ (vacuum energy)
\cite{Einstein:1917,cc} which has the equation of state $w=-1$.
However, as is well known, there are two difficulties arise from the
cosmological constant scenario, namely the two famous cosmological
constant problems --- the ``fine-tuning'' problem and the ``cosmic
coincidence'' problem \cite{coincidence}. The fine-tuning problem
asks why the vacuum energy density today is so small compared to
typical particle scales. The vacuum energy density is of order
$10^{-47} {\rm GeV}^4$, which appears to require the introduction of
a new mass scale 14 or so orders of magnitude smaller than the
electroweak scale. The second difficulty, the cosmic coincidence
problem, says: Since the energy densities of vacuum energy and dark
matter scale so differently during the expansion history of the
universe, why are they nearly equal today? To get this coincidence,
it appears that their ratio must be set to a specific, infinitesimal
value in the very early universe.

Theorists have made lots of efforts to try to resolve the
cosmological constant problem, but all these efforts were turned out
to be unsuccessful.\footnote{Of course the theoretical consideration
is still in process and has made some progresses. In recent years,
many string theorists have devoted to understand and shed light on
the cosmological constant or dark energy within the string
framework. The famous Kachru-Kallosh-Linde-Trivedi (KKLT) model
\cite{kklt} is a typical example, which tries to construct
metastable de Sitter vacua in the light of type IIB string theory.
Furthermore, string landscape idea \cite{landscape} has been
proposed for shedding light on the cosmological constant problem
based upon the anthropic principle and multiverse speculation.}
However, there remain other candidates to explaining dark energy. An
alternative proposal for dark energy is the dynamical dark energy
scenario. The cosmological constant puzzles may be better
interpreted by assuming that the vacuum energy is canceled to
exactly zero by some unknown mechanism and introducing a dark energy
component with a dynamically variable equation of state. The
dynamical dark energy proposal is often realized by some scalar
field mechanism which suggests that the energy form with negative
pressure is provided by a scalar field evolving down a proper
potential. Actually, this mechanism is enlightened to a great extent
by the inflationary cosmology. As we have known, the occurrence of
the current accelerating expansion of the universe is not the first
time in the expansion history of the universe. There is significant
observational evidence strongly supports that the universe underwent
an early inflationary epoch, over sufficiently small time scales,
during which its expansion rapidly accelerated under the driven of
an ``inflaton'' field which had properties similar to those of a
cosmological constant. The inflaton field, to some extent, can be
viewed as a kind of dynamically evolving dark energy. Hence, the
scalar field models involving a minimally coupled scalar field are
proposed, inspired by inflationary cosmology, to construct
dynamically evolving models of dark energy. The only difference
between the dynamical scalar-field dark energy and the inflaton is
the energy scale they possess. Famous examples of scalar-field dark
energy models include quintessence \cite{quintessence}, $K$-essence
\cite{kessence}, tachyon \cite{tachyon}, phantom \cite{phantom},
ghost condensate \cite{ghost1,ghost2} and quintom \cite{quintom},
and so forth. Generically, there are two points of view on the
scalar-field models of dynamical dark energy. One viewpoint regards
the scalar field as a fundamental field of the nature. The nature of
dark energy is, according to this viewpoint, completely attributed
to some fundamental scalar field which is omnipresent in
supersymmetric field theories and in string/M theory. The other
viewpoint supports that the scalar field model is an effective
description of an underlying theory of dark energy. In any case, the
scalar field dark energy models must face the test of cosmological
observations. A typical approach for this is to predict the
cosmological evolution behavior of the models, such as the evolution
of equation-of-state or Hubble parameter, by putting in the
Lagrangian (in particular the potential) by hand or theoretically,
and to make a consistency check of models by comparing it with
observations. An alternative approach is to start from observational
data and to reconstruct corresponding theoretical Lagrangian. It
looks that the latter is more efficient to find out the best-fit
models of dark energy from observations.

The accumulation of the current observational data has opened a
robust window for probing the recent dynamical behavior of dark
energy. An intriguing aspect in the study of dark energy is that the
analysis of the observational data of type Ia supernova, mainly the
157 ``gold'' data listed in Riess et al. \cite{Riess:2004nr}
including 14 high redshift data from the {\it Hubble Space
Telescope} (HST)/Great Observatories Origins Deep Survey (GOODS)
program and previous data, using the parametrization of Hubble
parameter $H(z)$ or equation-of-state of dark energy $w(z)$, shows
that the equation of state of dark energy $w$ is likely to cross the
cosmological-constant boundary $-1$ (or phantom divide), i.e. $w$ is
larger than $-1$ in the recent past and less than $-1$ today
\cite{Alam:2004jy,Huterer:2004ch,quintom}. The dynamical evolving
behavior of dark energy with $w$ getting across $-1$ has brought
forward great challenge to the model-building of scalar-field in the
cosmology. The conventional scalar-field model, the quintessence
with a canonical kinetic term, can only evolve in the region of
$w\geqslant -1$, whereas the model of phantom with negative kinetic
term can always lead to $w\leqslant -1$. Neither the quintessence
nor the phantom alone can realize the transition of $w$ from $w>-1$
to $w<-1$ or vice versa. Although the $K$-essence can realize both
$w>-1$ and $w<-1$, it has been shown that it is very difficult for
$K$-essence to achieve $w$ of crossing $-1$ \cite{Vikman:2004dc}.
Hence, the quintom model was proposed for describing the dynamical
evolving behavior of $w$ crossing $-1$ \cite{quintom}.

The nomenclature ``quintom'' is suggested in the sense that its
behavior resembles the combined behavior of quintessence and
phantom. Thus, a simple realization of quintom scenario is a model
with the double fields of quintessence and phantom \cite{quintom}.
The cosmological evolution of such model has been investigated in
detail \cite{twofield}. It should be noted that such a quintom model
would typically encounter the problem of quantum instability
inherited from the phantom component. For the single real scalar
field models,\footnote{It has been proposed that a noncanonical
model of single complex scalar field, called ``hessence'', can
successfully realize the quintom-like behavior of $w=-1$ crossing
\cite{Wei:2005nw}.} the transition of crossing $-1$ for $w$ can
occur for the Lagrangian density $p(\phi, X)$, where X is a
kinematic term of a scalar-field $\phi$, in which
$\partial{p}/\partial{X}$ changes sign from positive to negative,
thus we require nonlinear terms in $X$ to realize the $w=-1$
crossing \cite{ghost2,Vikman:2004dc,Anisimov:2005ne}. When adding a
high derivative term to the kinetic term $X$ in the single scalar
field model, the energy-momentum tensor is proven to be equivalent
to that of a two-field quintom model \cite{Li:2005fm}.
In addition, it is remarkable that the generalized ghost condensate
model of a single-real-scalar-field is a successful realization of
the quintom-like dark energy \cite{Tsujikawa:2005ju,Zhang:2006qu}.
In Ref.\cite{ghost2}, a dark energy model with a ghost scalar field
has been explored in the context of the runaway dilaton scenario in
low-energy effective string theory. The authors addressed for the
dilatonic ghost condensate model the problem of vacuum stability by
implementing higher-order derivative terms and showed that a
cosmological model of quintom-like dark energy can be constructed
without violating the stability of quantum fluctuations.
Furthermore, a generalized ghost condensate model was investigated
in Refs.\cite{Tsujikawa:2005ju,Zhang:2006qu} by means of the
cosmological reconstruction program. For another interesting
single-field quintom model see Ref.\cite{Huang:2005gu}, where the
$w=-1$ crossing is implemented with the help of a fixed background
vector field. Besides, there are also many other interesting models,
such as holographic dark energy model \cite{holo} and braneworld
model \cite{Cai:2005ie}, being able to realize the quintom-like
behavior. In this paper we will focus on the generalized ghost
condensate model and will reconstruct this model using various dark
energy parametrizations and the up-to-date observational datasets.

This paper is organized as follows: In section \ref{sec:para} we
address the various dark energy parametrizations and describe the
analysis results of the current experimental data of various
astronomical observations. In section \ref{sec:ghost} we perform a
cosmological reconstruction for the generalized ghost condensate
model from the dark energy parametrizations and the fitting results
of the up-to-date observational data. Finally we give the concluding
remarks in section \ref{sec:concl}.

\section{Dark energy parametrizations and results of observational
constraints}\label{sec:para}

The distinctive feature of the cosmological constant or vacuum
energy is that its equation of state is always exactly equal to
$-1$. Whereas, the dynamical dark energy exhibits a dynamic feature
that its equation-of-state as well as its energy density are
evolutionary with time. An efficient approach to probing the
dynamics of dark energy is to parameterize dark energy and then to
determine the parameters using various observational data. One can
explore the dynamical evolution behavior of dark energy efficiently
by making use of this way, although the results obtained are
dependent on the parametrizations of dark energy more or less. Among
the various parametric forms of dark energy, the minimum complexity
required to detect time variation in dark energy is to add a second
parameter to measure a change in the equation-of-state parameter
with redshift. This is the so-called linear expansion
parametrization $w(z)=w_0+w'z$, where $w'\equiv dw/dz|_{z=0}$, which
was first used by Di Pietro $\&$ Claeskens \cite{DiPietro:2002cz}
and later by Riess et al. \cite{Riess:2004nr}. However, when some
``longer-armed'' observations, e.g. CMB and LSS data, are taken into
account, this form of $w(z)$ will be unsuitable due to the
divergence at high redshift. A frequently used parametrization form
of equation-of-state, $w(z)=w_0+w_a z/(1+z)$, suggested by
Chevallier $\&$ Polarski \cite{Chevallier:2000qy} and Linder
\cite{Linder:2002et}, can avoid the divergence problem effectively.
It should be noted that this parametrization form has been
investigated enormously in exploring the dynamical property of dark
energy in the light of observational data. We shall summarize some
main constraint results for this parametrization in the follows.

A recently popular method of constraining the dark energy
parametrization is the so-called ``global fitting'' which tries to
make use of the most observational information including CMB, SNIa
and LSS data as well as to consider the dark energy perturbation,
employing the Markov chain Monte Carlo (MCMC) techniques. In the
global fitting one usually should determine an eight-dimensional set
of cosmological parameters, ${\bf P}\equiv (\omega_b, \omega_c,
\Theta_S, \tau, w_0, w_a, n_s, \log[10^{10} A_s])$, where
$\omega_b=\Omega_b h^2$ and $\omega_c=\Omega_c h^2$ are baryon and
cold dark matter densities relative to the critical density,
$\Theta_S$ is the ratio (multiplied by 100) of the sound horizon and
the angular diameter distance, $\tau$ is the optical depth, $A_s$ is
defined as the amplitude of the primordial scalar power spectra, and
$n_s$ measures the spectral index. In Ref.\cite{Xia:2005ge}, using
the first-year WMAP (Wilkinson Microwave Anisotropy Probe)
temperature and polarization data
\cite{Bennett:2003bz,Hinshaw:2003ex}, the 3D power spectrum data of
galaxies from the SDSS (Sloan Digital Sky Survey)
\cite{Tegmark:2003uf} and the ``gold'' dataset of 157 SNIa
\cite{Riess:2004nr}, the authors obtained the following fit results:
for with dark energy perturbation, $\Omega_{\rm
m0}=0.319^{+0.030}_{-0.031}$, $w_0=-1.167^{+0.191}_{-0.190}$ and
$w_a=0.597^{+0.657}_{-0.713}$; for without dark energy perturbation,
$\Omega_{\rm m0}=0.314^{+0.031}_{-0.031}$,
$w_0=-1.098^{+0.078}_{-0.080}$ and $w_a=0.416^{+0.293}_{-0.153}$.
Subsequently, with the announcement of the latest three-year WMAP
data \cite{wmap3}, Zhao et al. \cite{Zhao:2006bt} extended their
previous results in Ref.\cite{Xia:2005ge} and got the fit results
as\footnote{In this fitting the SDSS information includes the 3D
power spectrum of galaxies (SDSS-gal) \cite{Tegmark:2003uf} and the
Lyman-$\alpha$ forrest (SDSS-lya) \cite{McDonald:2004xn} data.}: for
with dark energy perturbation, $w_0=-1.146^{+0.176}_{-0.178}$ and
$w_a=0.600^{+0.622}_{-0.652}$; for without dark energy perturbation,
$w_0=-1.118^{+0.152}_{-0.147}$ and $w_a=0.499^{+0.453}_{-0.498}$.
Lately, Riess et al. \cite{Riess:2006fw} released the new ``gold''
dataset of 182 SNIa, in which the full sample of 23 SNIa at
$z\geqslant 1$ provides the highest-redshift sample known. In
addition, it should be mentioned that the gamma ray bursts (GRBs)
have been, in some studies, processed as ``known candles'' due to
some intrinsic correlations between temporal or spectral properties
of GRBs and their isotropic energies and luminosities. Such
investigations have triggered studies on using GRBs as cosmological
probes. The largest GRB sample compiled by Shaefer was released
lately \cite{Schaefer:2006pa}. In a recent work \cite{Li:2006ev},
for including the most observational information in the analysis of
probing the dynamical dark energy, the authors added the GRB data to
the global fitting.\footnote{The fitting analysis made use of the
data of three-year WMAP \cite{wmap3}, the 3D power spectrum of
galaxies from SDSS \cite{Tegmark:2003uf} and from 2dFGRS
\cite{Cole:2005sx}, and the 182 ``gold'' data of SNIa
\cite{Riess:2006fw}. And, the dark energy perturbation has been
considered in the fitting analysis.} Their fit results can be
summarized as: for with GRB data, $\Omega_{\rm
m0}=0.296^{+0.023}_{-0.019}$, $w_0=-1.09^{+0.22}_{-0.06}$ and
$w_a=0.89^{+0.11}_{-0.74}$; for without GRB data, $\Omega_{\rm
m0}=0.300^{+0.009}_{-0.033}$, $w_0=-1.09^{+0.27}_{-0.06}$ and
$w_a=0.90^{+0.07}_{-1.01}$.

An alternative approach to constraining the dynamical dark energy
parametrization is to use the measured value of the CMB shift
parameter, together with the baryon acoustic oscillation (BAO)
measurement from the SDSS, and the SNIa data, which provides a more
economical scheme comparing to the global fitting method. The CMB
shift parameter $R$ is perhaps the least model-dependent parameter
that can be extracted from CMB data, since it is independent of
$H_0$. The shift parameter $R$ is given by \cite{Bond:1997wr}
$R\equiv \Omega_{\rm m0}^{1/2}\int_0^{z_{\rm CMB}}dz'/E(z')$, where
$z_{\rm CMB}=1089$ is the redshift of recombination and $E(z)\equiv
H(z)/H_0$. The value of the shift parameter $R$ can be determined by
three-year integrated WMAP analysis \cite{wmap3}, and has been
updated by Wand $\&$ Mukherjee \cite{Wang:2006ts} to be $1.70\pm
0.03$ independent of the dark energy model. The measurement of the
BAO peak in the distribution of SDSS luminous red galaxies (LRGs)
\cite{Eisenstein:2005su} gives $A=0.469(n_s/0.98)^{-0.35}\pm 0.017$
(independent of a dark energy model) at $z_{\rm BAO}=0.35$, where
$A$ is defined as $A\equiv \Omega_{\rm m0}^{1/2} E(z_{\rm
BAO})^{-1/3}[(1/z_{\rm BAO})\int_0^{z_{\rm BAO}}dz'/E(z')]^{2/3}$.
Here the scalar spectral index is taken to be $n_s=0.95$ as measured
by the three-year WMAP data \cite{wmap3}. Wang and Mukherjee
\cite{Wang:2006ts} used this method\footnote{In this analysis, the
authors used the SNIa data from the HST/GOODS program
\cite{Riess:2004nr},157 ``gold'' data, labeled as ``Riess04'', and
the first-year Supernova Legacy Survey (SNLS) \cite{Astier:2005qq},
labeled as ``Astier05''.} and found the constraints on the Linder
parametrization: for Riess04+WMAP3+SDSS,
$w_0=-0.813^{+0.293}_{-0.296}$ and $w_a=-0.510^{+1.265}_{-1.259}$;
for Astier05+WMAP3+SDSS, $w_0=-1.017^{+0.199}_{-0.200}$ and
$w_a=-0.039^{+1.045}_{-1.052}$. Such results are qualitatively
consistent with those of Zhao et al. \cite{Zhao:2006bt}, with
significant differences that may be explained by the differences in
the combination of data used, and perhaps by some data analysis
details.

The aforementioned discussions are focussed on the Linder
parametrization of dark energy for the equation of state. Besides
the Linder parametrization, there are also some other
parametrization forms for the dark energy equation-of-state
parameter, for instance, the form of $w(z)=w_0+w_b z/(1+z)^2$
\cite{Jassal:2004ej}. In addition, the parametrization for the
Hubble parameter $H(z)$ is also often considered. The form can be
expressed as\footnote{The corresponding equation-of-state of dark
energy is $w(x)=-1+{A_1x+2A_2x^2\over 3(A_0+A_1x+A_2x^2)}$, where
$x=1+z$.} \cite{Alam:2003fg} $E(z)=[\Omega_{\rm
m0}(1+z)^3+A_0+A_1(1+z)+A_2(1+z)^2]^{1/2}$, where
$A_0+A_1+A_2=1-\Omega_{\rm m0}$. This form is an interpolating fit
for $E^2(z)$ having the right behavior for both small and large
redshifts. In this paper, for providing the base for reconstruction
of the scalar-field dark energy model, we shall consider these three
parametrization forms for dark energy: $w(z)=w_0+w_a z/(1+z)$,
marked as ``parametrization 1''; $w(z)=w_0+w_b z/(1+z)^2$, marked as
``parametrization 2''; $E(z)=[\Omega_{\rm
m0}(1+z)^3+A_0+A_1(1+z)+A_2(1+z)^2]^{1/2}$, marked as
``parametrization 3''.

\begin{figure}[htbp]
\begin{center}
\includegraphics[scale=0.90]{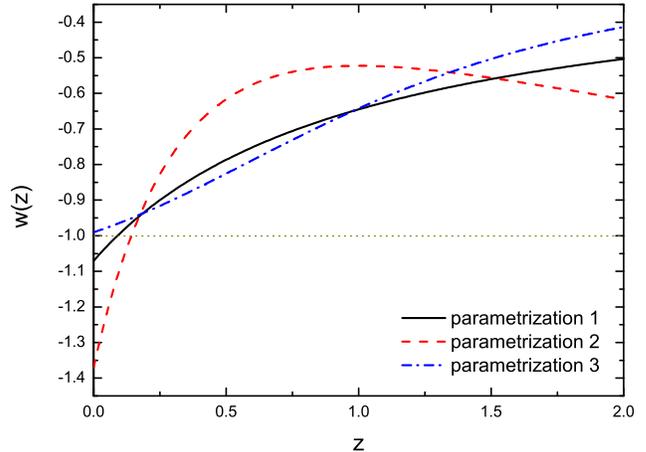}
\caption[]{\small The evolutions of the equation of state of dark
energy, corresponding to three forms of parametrization,
respectively. Parametrization 1 is $w(z)=w_0+w_a z/(1+z)$,
parametrization 2 is $w(z)=w_0+w_b z/(1+z)^2$ and parametrization 3
is $E(z)=[\Omega_{\rm m0}(1+z)^3+A_0+A_1(1+z)+A_2(1+z)^2]^{1/2}$.
Here we use the best-fit values of the joint analysis of
SNIa+CMB+LSS \cite{Gong:2006gs}. In the concrete, for
parametrization 1, $\Omega_{\rm m0}=0.29$, $w_0=-1.07$ and
$w_a=0.85$; for parametrization 2, $\Omega_{\rm m0}=0.28$,
$w_0=-1.37$ and $w_b=3.39$; for parametrization 3, $\Omega_{\rm
m0}=0.30$, $A_1=-0.48$ and $A_2=0.25$.}\label{fig:eoswz}
\end{center}
\end{figure}

These three forms of parametrization have lately been considered by
Gong $\&$ Wang \cite{Gong:2006gs}. In addition, it should be
mentioned that similar results were also independently obtained in
Ref.\cite{Nesseris:2006ey} for the ``parametrization 1'' and in
Ref.\cite{Alam:2006kj} for the ``parametrization 3''. In
Ref.\cite{Gong:2006gs}, the authors constrained the parameters using
the new measurement of the CMB shift parameter \cite{Wang:2006ts},
together with LSS data (the BAO measurement from the SDSS LRGs)
\cite{Eisenstein:2005su} and SNIa data (182 ``gold'' data released
recently) \cite{Riess:2006fw}. The fit results are summarized as
follows: for ``parametrization 1'', $\Omega_{\rm m0}=0.29\pm 0.04$,
$w_0=-1.07^{+0.33}_{-0.28}$ and $w_a=0.85^{+0.61}_{-1.38}$; for
``parametrization 2'', $\Omega_{\rm m0}=0.28^{+0.04}_{-0.03}$,
$w_0=-1.37^{+0.58}_{-0.57}$ and $w_b=3.39^{+3.51}_{-3.93}$; for
``parametrization 3'', $\Omega_{\rm m0}=0.30\pm 0.04$,
$A_1=-0.48^{+1.36}_{-1.47}$ and $A_2=0.25^{+0.52}_{-0.45}$. The
three forms of parametrization are investigated uniformly in this
work, so it is convenient to compare them with each other. The cases
for evolution of $w(z)$ are plotted in Fig.\ref{fig:eoswz}, using
the best-fit results. We shall use these fit results to reconstruct
the generalized ghost condensate model in the next section.

\section{Generalized ghost condensate model and its
reconstruction}\label{sec:ghost}

The reconstruction of scalar-field dark energy models has been
widely studied. For a minimally coupled scalar field with a
potential $V(\phi)$, the reconstruction is simple and
straightforward \cite{simplescalar}. Saini et al.
\cite{Saini:1999ba} reconstructed the potential and the equation of
state of the quintessence field by parameterizing the Hubble
parameter $H(z)$ based on a versatile analytical form of the
luminosity distance $d_L(z)$. This method can be generalized to a
variety of models, such as scalar-tensor theories
\cite{scalartensor}, $f(R)$ gravity \cite{frgrav}, $K$-essence model
\cite{Li:2006bx}, and also hessence model \cite{Zhao:2006mp}, etc..
For the reconstruction of general scalar-field dark energy models,
Tsujikawa has investigated in detail \cite{Tsujikawa:2005ju}. In
this section, we shall focuss on the reconstruction of the
generalized ghost condensate model based on the aforementioned three
forms of parametrization.

First, let us consider the Lagrangian density of a general scalar
field $p(\phi, X)$, where
$X=-g^{\mu\nu}\partial_\mu\phi\partial_\nu\phi/2$ is the kinetic
energy term. Note that $p(\phi, X)$ is a general function of $\phi$
and $X$, and we have used a sign notation $(-, +, +, +)$.
Identifying the energy momentum tensor of the scalar field with that
of a perfect fluid, we can easily derive the energy density of dark
energy, $\rho_{\rm de}=2Xp_X-p$, where $p_X=\partial p/\partial X$.
Thus, in a spatially flat Friedmann-Robertson-Walker (FRW) universe
involving dust matter (baryon plus dark matter) and dark energy, the
dynamic equations for the scalar field are
\begin{equation}
3H^2=\rho_{\rm m}+2Xp_X-p,\label{hsqr}
\end{equation}
\begin{equation}
2\dot{H}=-\rho_{\rm m}-2Xp_X,\label{hdot}
\end{equation}
where $X=\dot{\phi}^2/2$ in the cosmological context, and note that
we have used the unit\footnote{This is the unit of Planck
normalization, here $M_P\equiv 1/\sqrt{8\pi G}$ is the reduced
Planck mass.} $M_P=1$ for convenience. Introducing a dimensionless
quantity
\begin{equation}
r\equiv E^2= H^2/H_0^2,
\end{equation}
we find from Eqs.(\ref{hsqr}) and (\ref{hdot}) that
\begin{equation}
p=[(1+z)r'-3r]H_0^2,\label{p}
\end{equation}
\begin{equation}
\phi'^2p_X={r'-3\Omega_{\rm m0}(1+z)^2\over r(1+z)},\label{px}
\end{equation}
where prime denotes a derivative with respect to $z$. The equation
of state for dark energy is given by
\begin{equation}
w={p\over \dot{\phi}^2 p_X-p}={(1+z)r'-3r\over 3r-3\Omega_{\rm
m0}(1+z)^3}.
\end{equation}
Next, let us consider the generalized ghost condensate model
proposed in Ref.\cite{Tsujikawa:2005ju} (see also
Ref.\cite{Zhang:2006qu}), in which the behavior of crossing the
cosmological-constant boundary can be realized, with the Lagrangian
density
\begin{equation}
p=-X+h(\phi)X^2,
\end{equation}
where $h(\phi)$ is a function in terms of $\phi$. Dilatonic ghost
condensate model \cite{ghost2} corresponds to a choice
$h(\phi)=ce^{\lambda\phi}$. From Eqs. (\ref{p}) and (\ref{px}) we
obtain
\begin{equation}
\phi'^2={12r-3(1+z)r'-3\Omega_{\rm m0}(1+z)^3\over r(1+z)^2},
\label{phip}
\end{equation}
\begin{equation}
h(\phi)={6(2(1+z)r'-6r+r(1+z)^2\phi'^2)\over
r^2(1+z)^4\phi'^4}\rho_{\rm c0}^{-1}, \label{hz}
\end{equation}
where $\rho_{\rm c0}=3H_0^2$ represents the present critical density
of the universe. The evolution of the field $\phi$ can be derived by
integrating $\phi'$ according to Eq.(\ref{phip}). Note that the
field $\phi$ is determined up to an additive constant $\phi_0$, so
it is convenient to take $\phi$ to be zero at the present epoch
($z=0$). The function $h(\phi)$ can be reconstructed using
Eq.(\ref{hz}) when the information of $r(z)$ is obtained from the
observational data.

\begin{figure}[htbp]
\begin{center}
\includegraphics[scale=0.9]{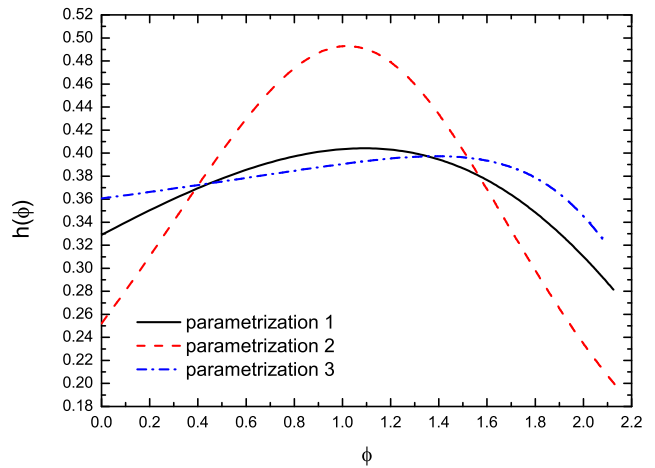}
\caption[]{\small Reconstruction of the generalized ghost condensate
model according to three forms of parametrization for dynamical dark
energy with the best-fit values of the parameters. Parametrization 1
is $w(z)=w_0+w_a z/(1+z)$, parametrization 2 is $w(z)=w_0+w_b
z/(1+z)^2$ and parametrization 3 is $E(z)=[\Omega_{\rm
m0}(1+z)^3+A_0+A_1(1+z)+A_2(1+z)^2]^{1/2}$. In this plot, we show
the cases of function $h(\phi)$, in unit of $\rho_{\rm c0}^{-1}$,
corresponding to the best-fit results of the joint analysis of
SNIa+CMB+LSS.}\label{fig:hphi}
\end{center}
\end{figure}

\begin{figure}[htbp]
\begin{center}
\includegraphics[scale=0.9]{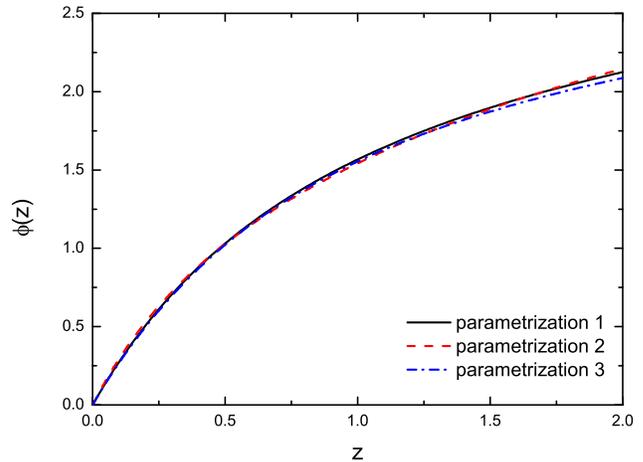}
\caption[]{\small Reconstruction of the generalized ghost condensate
model according to three forms of parametrization for dynamical dark
energy with the best-fit values of the parameters. Parametrization 1
is $w(z)=w_0+w_a z/(1+z)$, parametrization 2 is $w(z)=w_0+w_b
z/(1+z)^2$ and parametrization 3 is $E(z)=[\Omega_{\rm
m0}(1+z)^3+A_0+A_1(1+z)+A_2(1+z)^2]^{1/2}$. In this plot, we show
the evolutions of the scalar field $\phi(z)$, in unit of the Planck
mass $M_P$ (note that here the Planck normalization $M_P=1$ has been
used), corresponding to the best-fit results of the joint analysis
of SNIa+CMB+LSS.}\label{fig:phiz}
\end{center}
\end{figure}

Generically, the Friedmann equation can be expressed as
\begin{equation}
r(z)=\Omega_{\rm m0}(1+z)^3+(1-\Omega_{\rm m0})f(z),
\end{equation}
where $f(z)$ is some function encoding the information about the
dynamical property of dark energy. For example, the
``parametrization 1'' corresponds to
$f(z)=(1+z)^{3(1+w_0+w_a)}\exp{[-3w_az/(1+z)]}$, and the
``parametrization 2'' corresponds to
$f(z)=(1+z)^{3(1+w_0)}\exp{[3w_bz^2/2(1+z)^2]}$. Whereas, for the
``parametrization 3'' one can straightforwardly write
$r(z)=\Omega_{\rm m0}(1+z)^3+A_0+A_1(1+z)+A_2(1+z)^2$. Hence, we can
reconstruct the function $h(\phi)$ for the generalized ghost
condensate model in the light of these forms of parametrization and
the corresponding fit results of the observational constraints. The
reconstruction for $h(\phi)$ is plotted in Fig.\ref{fig:hphi}, using
the three parametrizations and the corresponding best-fit values of
parameters from the observational data analysis of SNIa+CMB+LSS. The
differences between the shapes of $h(\phi)$ comes from the
differences in these forms of parametrization because that
uncertainties still remain large for ``model-independent''
observational constrains of dynamical dark energy. The crossing of
the cosmological-constant boundary corresponds to $hX=1/2$. The
system can enter the phantom region ($hX <1/2$) without
discontinuous behavior of $h$ and $X$. In addition, the evolution of
the scalar field $\phi(z)$ is also determined by the reconstruction
program, see Fig.\ref{fig:phiz}. We see that the differences in the
shapes of $\phi(z)$ are very little. It should be mentioned that the
reconstruction of the generalized ghost condensate model has been
carried out in Ref.\cite{Tsujikawa:2005ju} in the light of
``parametrization 3'' from the best-fit results of the SNIa gold
dataset \cite{Riess:2004nr}. However, the constraints from only the
157 gold data of SNIa can only give very preliminary results
\cite{Lazkoz:2005sp}: $A_1=-4.16\pm 2.53$ and $A_2=1.67\pm 1.03$,
for the prior $\Omega_{\rm m0}=0.3$. It can be seen clearly that
these results have significant differences from those derived from
new observational data analysis. So our result of reconstruction
significantly improves the previous result.

In addition, as has been pointed out by Tsujikawa
\cite{Tsujikawa:2005ju}, it should be cautioned that the
perturbation of the field $\phi$ is plagued by a quantum instability
whenever it behaves as a phantom \cite{ghost2}. Even at the
classical level the perturbation becomes unstable for $1/6<hX<1/2$,
because that the speed of sound, $c_s^2=p_X/(p_X+2Xp_{XX})$, will
become negative. This instability may be avoided if the phantom
behavior is just transient. In fact the dilatonic ghost condensate
model can realize a transient phantom behavior (see, e.g., Fig.4 in
Ref.\cite{ghost2}). In this case the cosmological-constant boundary
crossing occurs again in the future, after which the perturbation
will become stable. Nevertheless, one may argue that the field can
be regarded as an effective one so as to evade problems such as
stability. In particular, the present focus is that how to establish
a dynamical scalar-field model on phenomenological level to describe
the possible dynamics of dark energy observed, disregarding the
field is fundamental or not.

\section{Concluding remarks}\label{sec:concl}

In this paper, we have reconstructed the ghost condensate
scalar-field model in the light of three forms of parametrization
for dynamical dark energy with fit results of observational data for
the parameters. The analysis of the data of astronomical
observations suggests that dark energy may possess dynamical nature,
i.e. the energy density as well as the equation of state are likely
to exhibit dynamical evolution property during the expansion history
of the universe. Furthermore, it is intriguing that recent analysis
of observational data, especially the ``gold'' SNIa data, shows that
the equation-of-state parameter of dark energy may, with some
possibility, cross the cosmological-constant boundary $w=-1$ during
the evolution of the universe. Although the scalar-field models of
dark energy, such as quintessence and phantom, can provide us with
dynamical mechanism for dark energy, the behavior of
cosmological-constant crossing brings forward a great challenge to
the model-building for dynamical dark energy, because neither
quintessence nor phantom can realize this manner. The model of
quintom was suggested to realize this behavior by means of the
incorporation of the features of quintessence and phantom. The
generalized ghost condensate model provides us with a successful
single-real-scalar-field model for realizing the quintom-like
behavior. For probing the dynamical nature of dark energy, one
should parameterize dark energy first and then constrain the
parameters using the observational data. In this paper, we reviewed
the recent constraint results for various parametrization forms.
Based upon three forms of parametrization for dynamical dark energy,
$w(z)=w_0+w_a z/(1+z)$, $w(z)=w_0+w_b z/(1+z)^2$ and
$E(z)=[\Omega_{\rm m0}(1+z)^3+A_0+A_1(1+z)+A_2(1+z)^2]^{1/2}$, with
the best-fit values of parameters, we perform a reconstruction for
the generalized ghost condensate model. The results of
reconstruction show that there are some differences in the various
forms of parametrization.

On the other hand, it should be noted that the cosmological constant
will be more favored than a dynamical dark energy when the SNLS
supernova dataset is considered instead of the ``gold'' one (see
e.g. Ref.\cite{Alam:2006kj} for details). This statement has become
even stronger after the recent appearance of the ESSENCE dataset
\cite{Wood-Vasey:2007jb,Davis:2007na}. However, though the
cosmological constant receives support from the SNLS+ESSENCE
dataset, the dynamical dark energy can not be ruled out yet.
Actually, it is difficult to reach firm conclusion on the property
of dark energy from these data until strong model-independent
analysis can be carried out. The increase of the quantity and
quality of observational data in the future will undoubtedly provide
a true {\it model-independent} manner for exploring the property of
dark energy. We hope that the future high-precision observations
(e.g. SNAP) may be capable of providing us with deep insight into
the nature of dark energy driving the acceleration of the universe.

\section*{ACKNOWLEDGMENTS}


It is a pleasure for us to thank Miao Li for useful discussions and
for reading the draft of this paper. This work is supported in part
by the Natural Science Foundation of China. XZ also thanks the
financial support from China Postdoctoral Science Foundation.


\end{document}